\documentclass[preprint,12pt]{elsarticle}

\usepackage{amssymb}
\usepackage{subcaption}
\usepackage{amssymb}
\usepackage{amsmath, amsfonts}
\usepackage{dsfont}
\usepackage{pifont}
\usepackage{lipsum}
\usepackage{float}
\usepackage{enumitem}   
\usepackage{graphicx}
\usepackage{epstopdf}
\usepackage{multirow}
\usepackage{makecell}
\usepackage{ragged2e}
\usepackage{tabularx}
\usepackage{relsize}
\usepackage{array}

\graphicspath{{./figures/}}

\begin{document}

\begin{frontmatter}



\title{Critical Node Detection in Temporal Social Networks, Based on Global and Semi-local Centrality Measures}


\affiliation[inst1]{organization={Algorithms and Computations},
            addressline={University of Tehran}, 
            city={Tehran},
            country={Iran}}
\affiliation[inst2]{organization={Department of Economics},
            addressline={Ghent University}, 
            city={Ghent},
            country={Belgium}}
            
\affiliation[inst3]{organization={Department of Physics and Astronomy},
            addressline={Ghent University}, 
            city={Ghent},
            country={Belgium}}
\author[inst1]{Zahra Farahi}
\author[inst1]{Ali Kamandi}
\author[inst1]{Rooholah Abedian}
\author[inst2,inst3]{Luis E C Rocha}

\begin{abstract}
Nodes that play strategic roles in networks are called critical or influential nodes. For example, in an epidemic, we can control the infection spread by isolating critical nodes; in marketing, we can use certain nodes as the initial spreaders aiming to reach the largest part of the network, or they can be selected for removal in targeted attacks to maximise the fragmentation of the network. In this study, we focus on critical node detection in temporal networks. We propose three new measures to identify the critical nodes in temporal networks: the temporal supracycle ratio, temporal semi-local integration, and temporal semi-local centrality. We analyse the performance of these measures based on their effect on the SIR epidemic model in three scenarios: isolating the influential nodes when an epidemic happens, using the influential nodes as seeds of the epidemic, or removing them to analyse the robustness of the network. We compare the results with existing centrality measures, particularly temporal betweenness, temporal centrality, and temporal degree deviation. The results show that the introduced measures help identify influential nodes more accurately. The proposed methods can be used to detect nodes that need to be isolated to reduce the spread of an epidemic or as initial nodes to speedup dissemination of information.
\end{abstract}


\begin{highlights}
\item critical Node Detection in Temporal Networks: identifying influential nodes in temporal networks is crucial for controlling phenomena like epidemics and network attacks.

\item Novel Measures Proposed: Three new measures are introduced to assess node importance in temporal networks.

\item Impact on Epidemic Spread and Network Robustness: Analyses of the effects of these measures on SIR epidemics and network robustness, highlighting their potential applications in regulating the spread and network resilience.

\item High Agreement Among Measures: Comparison with existing measures indicates that the novel measures achieve higher performance in identifying critical nodes in temporal networks.
\end{highlights}


\end{frontmatter}



\section{Introduction}
\label{sec:intro}

Network science is a useful and appropriate mathematical framework to represent social contacts and relations between objects in the real world. Network science is used in various disciplines, from biological to social sciences. Network models have been applied to study protein-protein interaction networks~ \cite{26burke2023towards,30goos2022human}, brain networks \cite{27luo2022survey} and neurosciences \cite{38vavsa2022null}, social networks \cite{32wang2022rumor, 76zarei2024bursts, 52cattuto2010dynamics, 54mastrandrea2015contact}, transport network and routing problems \cite{21-ma2023improved}, the dynamics of epidemic \cite{25zhang2022interaction,48demongeot2020si,49han2021dynamical, Rocha2020}, non-fungible tokens (NFTs) \cite{39vasan2022quantifying}, the spread of violence within the social network of youngsters~\cite{GEERAERT2024102259} and, in combination with maximum entropy network models, misinformation campaigns on Twitter \cite{77de2022maximum}. The goal is to unveil how specific patterns of connections regulate or influence spreading process.

The complexity of real-world networks has prompted scientists to study different features of networks to unveil what topology they have and what behaviours they may show. Therefore, they proposed different measures to help analyse complex networks, such as the shortest path \cite{61cherkassky1996shortest}, betweennness \cite{66freeman1977set}, node or edge centrality \cite{60song2022important}, cycle and circuits, and robustness. Novel metrics based on existing fundamental metrics have also been introduced. For example, the Locality-based Structure System (LSS) is based on three different parameters: degree, nodes' \textit{k}-shell, and the number of triangles a node is involved \cite{67ullah2023lss}. The cluster coefficient ranking measure (ECRM) is based on the common hierarchy of a node and its neighbours \cite{64zareie2020finding}. Each node is labelled based on \textit{k}-shell algorithm, and the method uses the common labels in the hierarchy of a node and its neighbours. Another study proposed a generalised degree decomposition (GDD) algorithm to improve the drawbacks of the k-shell algorithm for critical node detection  \cite{73zheng2023new}. 

Node centrality measures help in finding the key or influential nodes. These nodes play a significant role in the network and provide network feedback to any changes in the network state. Different algorithms have been proposed to detect influential nodes: statistical-based, neural network-based, and diffusion-based, among others \cite{59ou2022identifying}. Depending on the network, the nodes represent different objects like humans, for example, in ref. \cite{70ahmed2023evaluating}, the author analyses the influence of nodes, representing researchers, using various metrics and proposes a comprehensive study of metrics to help researchers in different fields. Laplacian Distance is another method for analysing the node importance in complex networks, and the distance Laplacian centrality (DLC) can be used for critical node detection \cite{71yin2024identifying}. This centrality focuses on the node's role on a global scale using the graph energy. TempoRank, based on a random walk, is introduced for detecting the critical nodes in temporal networks \cite{78rocha2014random}.

Nosirov et al. \cite{56nosirov2022review} compiled diverse algorithms for determining the shortest path in networks and proposed a comprehensive classification for them. The significance of this measure becomes evident when researchers across different disciplines utilise it; for example, a model is proposed for message passing in neural networks, wherein each node propagates information to all its neighbours via the shortest path \cite{57abboud2022shortest}. Node degree, inverse local clustering coefficient (ILCC), and neighbours’ degree have been used to identify the most influential nodes \cite{65berahmand2019new}. A new degree centrality measure based on the spanning tree, called Multi-Spanning Tree-based Degree Centrality (MSTDC), was introduced for detecting the most influential nodes \cite{72dai2023identifying}. 

Influential nodes affect the spread of rumours, information, and epidemics. Researchers develop methods to detect the most influential or critical nodes, such as information diffusion in complex networks based on the SIS (Susceptible-Infected-Susceptible) epidemic model and information competition and cooperation \cite{25zhang2022interaction}. The robustness of the network is also an important object of study, especially under network attacks or failures. Robustness represents network strength against loss of nodes. Different measures for analysing robustness have been proposed, including the size of the largest connected component, entropy, strength, and skewness \cite{58oehlers2021graph}.

VoteRank-based methods have also been used to detect influential nodes \cite{69liu2021identifying}. In this method, in each turn, all nodes vote for their neighbours, and at the end of the turn, the node with the highest score is selected as one of the most influential nodes \cite{62zhang2016identifying}. Since the voting ability of nodes may be different and based on the coreness of the neighbours, an alternative is to use a coreness-based VoteRank called NCVoteRank \cite{63kumar2020identifying}. The Recent and Weight strategies have also been proposed to identify critical nodes in temporal networks for effective epidemic control by leveraging information about past temporal contact patterns \cite{79lee2012exploiting}. The IM-ELPR algorithm for critical node detection uses the \textit{h}-index to find the seed nodes \cite{68kumar2021elpr}. After detecting the network communities, it consolidates the small communities to achieve the larger ones and finds the \textit{k} most influential nodes. 

Several real-world networks such as the brain functional connectivity \cite{18ding2020towards}, fraud detection in banking \cite{19hajdu2020temporal}, epidemics like Covid-19 \cite{35yang2022complex,36myall2022prediction} or sexual infections \cite{75frieswijk2023time}, and the behaviour of mobile phone users \cite{34ventura2022epidemic} can be described by temporal networks. The ubiquity of temporal networks in representing the real world motivates researchers to focus on analysing their features and the impact of changing structures on dynamic processes, such as epidemic, information, and flow. 

Researchers have analysed the role of important nodes in information flow using centrality metrics such as betweenness and closeness \cite{8tang2010analysing, 31salama2022temporal}. A temporal walk centrality was proposed to analyse information flow \cite{11oettershagen2022temporal}. This algorithm is based on a temporal random walk to capture that diffusion spreads not only through the shortest path but can also be distributed to adjacency paths. 

Sampling-based algorithms for temporal betweenness \cite{3cruciani2023approximating} and a method to find temporal paths in temporal networks considering waiting time \cite{9casteigts2021finding} have been proposed. The fact that people may forget about news and not continue to propagate information to others (memory or expiration time) has been used in an algorithm for the temporal reachable set \cite{16ban2022semi}. Apart from node centrality, the researchers also proposed models to measure edge centrality to identify the most important connections in the network \cite{16ban2022semi}. Other researchers have expanded the temporal network to a spatial-temporal network, a layered network made of networks, for example, ESTNet, for analysing and controlling traffic \cite{20luo2022estnet}. 

This study focuses on local, semi-local and global node centrality measures and introduce 3 novel metrics to measure the importance of nodes in temporal networks. We compare the results with popular centrality measures such as betweenness centrality, degree deviation, and closeness. In section \ref{methods}, we define temporal network and centrality measures, describe the data sets to be used, and the epidemic model used for performance testing. In section \ref{novel}, we introduce the proposed new metrics. In section~\ref{results}, we analyse and compare the proposed and existing metrics. Finally, section~\ref{conclusion} briefly discusses the proposed metrics and their accuracy.

\section{Materials and Methods}
\label{methods}

\subsection{Temporal networks}
 
A static network is defined as $G(N, E)$, where $N$ is a set of nodes and $E$ is a set of edges. In temporal networks, edges may be active and inactive from time to time, unlike a static network in which they are always active. For the temporal networks in the time interval $[t_s,t_e]$, we have $\mathcal{G} = (\mathcal{N},\mathcal{E},\mathcal{L})$ \cite{40zhan2020susceptible,41yang2023scalable}, where:

\begin{itemize}
  \item $\mathcal{N}$ is the set of nodes
  \item $\mathcal{E} = \{(i,j,t): ~ i~,~j \in (\mathcal{N})~\& ~t \in [t_s,t_e] \}$ is a set of edges active at time t.
  \item
       $\mathcal{L}_{(i,j,t)}= \begin{cases}
                       1, & \mbox{ } (i,j,t)\in \mathcal{E}, \\
                       0, & \mbox{otherwise}.
                     \end{cases}$
\end{itemize}

$\mathcal{G}$ comprises several snapshots of a graph, one per time step. The edges of the temporal network evolve, and one snapshot of the network can differ from the other at different times. The identical static network is the union of all $\mathcal{G}$ snapshots.

In the time interval $[t_{s}, t_e]$, the temporal degree of a node $i \in \mathcal{N}$ is the number of nodes $j \in \mathcal{N}$ are connected to $i$ in the time interval $[t_s,t_e]$ \cite{14zhang2022significant}:

\begin{equation}\label{temporal_degree_eq}
 d_{i,\mathcal{G}}(t_s,t_e) = \sum\limits_{u= t_s}^ {t_e}{\mathcal{L}(i,j,u)}.
\end{equation}

The connection between $i$ and $j$ may disconnect several times in the interval $[t_s, t_e]$, but if in one of the snapshots, $i$ and $j$ are connected, then we consider $j$ as a neighbour of $i$.

Node $j$ is the neighbour of node $i$ in $\mathcal{G}_{[t_S,t_e]}$ if and only if $\exists t \in [t_s,t_e] \& (i,j,t)\in \mathcal{E}$. Therefore, the set of all neighbors of $i$ in $[t_S,t_e]$ is:
\begin{equation}\label{temporal_neigh_eq}
  \Gamma_{i,\mathcal{G}_{(t_S,t_e)}} = \biggl\{v: v\in \mathcal{N} \; \& \; (i,j,t)\in \mathcal{E}, t \in [t_S,t_e]\biggr\}.
\end{equation}

A path in a static network is a sequence of edges connecting two nodes. The distance between two nodes is the number of edges in the path from the source node to the destination node. In temporal networks, a temporal path is a sequence of edges in $\mathcal{G}$ appearing in a sequence of time snapshots where every two consecutive edges have a common node. A sequence of $\mathcal{E}(i,j,t_0), \mathcal{E}(i,j,t_1),..., \mathcal{E}(i,j,t_n)$ represents the path between nodes $i$ and $j$. In other words, a temporal path is a sequence of edges $\bigl\{e_n(i_n,j_n,t_n,\delta{t_n}): n \in \mathds{N} \bigr\}$ which $i_n, j_n$ are in contact in $[t_n, \delta t_n]$ and edge $e_{n+1}$ appears in the path after edge $e_n$ if and only if one of the following conditions applies to it:

\begin{itemize}
  \item $t_n < t_{n+1}$
or
  \item $t_n > t_{n+1} \; \& \; \delta t_n < \delta t_{n+1}$
  \end{itemize}

\begin{equation}\label{temporal_path_eq}
  \mathcal{P}_{\mathcal{G},(i,j)} = \begin{cases}
                                     1, & \mbox{ $\exists$ ${ \mathcal{E}(i,j,t_k)}$, } \\
                                     0, & \mbox{otherwise}.
                                   \end{cases}
\end{equation}

The distance in the temporal network is the total number of time steps a node needs to reach the destination node from the source node. Consider a path that starts from node $j$ at time step $t$, after visiting a sequence of edges, reaches node $i$ at time step $t'$ \cite{42pan2011path,43holme2012temporal}, then:
\begin{equation}\label{temporal_distance_eq}
  \tau_{\mathcal{G},(u,v)} = \begin{cases}
                               t - t', & \mbox{ } \mathcal{P}_{\mathcal{G},(u,v)} = 1, \\
                               \infty, & \mbox{otherwise}.
                             \end{cases}
\end{equation}

A cycle is a path where the first and last nodes are the same. A temporal cycle is a temporal path in which the first and last nodes are the same. \textit{Base cycles} are the set of smallest cycles that make up the network. Therefore, they do not have any other node-induced sub-graph that makes a cycle. If $\mathbb{T}$ is the temporal spanning tree of $\mathcal{G}$ and $e(i,k,t) \in \mathcal{E}$ does not exist in $\mathcal{G}$ then $\mathbb{T} + e$ makes a base cycle of $\mathcal{G}$.Therefore, if $\mathcal{P}_{\mathcal{G},(i,k)} = \langle (i_0,i_1,t_0)\in\mathcal{E}, (i_1,i_2,t_1)\in\mathcal{E}, ... , (i_i,k,t_i)\in\mathcal{E} \rangle$ then:

\begin{equation}\label{eq:temporal_cycle}
  \mathcal{C}_\mathcal{G} = \bigcup{\biggl\{\mathcal{P}_{\mathcal{G},(i,k)}+(k,i,t)\in\mathcal{E}| (k,i,t) \notin \mathbb{T}, t> t_n \biggr\}}.
\end{equation}

\subsection{Centrality Measures}

There are three types of centrality measures: one considers the centrality of nodes locally in their neighbourhood, the second considers the node's importance on a global scale in the whole network, and the third is between the local and global indexes (hereafter called semi-local).

\textbf{Local Temporal Degree Centrality:} It only considers a node and its direct neighbourhood. It shows the number of nodes a node $i$ can affect directly \cite{13ye2023vital}. In temporal networks, the degree centrality of a node can be different at each time. The temporal degree centrality of node $j$ at time $t$ is the number of nodes connected to $j$ at time $t$ \cite{14zhang2022significant}.

\textbf{Local Temporal Degree Deviation (TDD):} It quantifies the difference between the temporal and static degrees. A higher value means that the contacts of a node are not always active because if TDD is small, the degree of the node is more or less constant over time. A higher temporal degree means that the node has an active connection in more time steps; therefore, it is more important in transferring a flow \cite{51yu2020identifying}.

\begin{equation}\label{eq:TDD}
  TDD(i) = \sqrt{\frac{1}{T}\sum\limits_{t=1}^ {T}{(D_t(i)-D(i))^2}}.
\end{equation}

\textbf{Semi-local centrality:} It considers not only the direct neighbours of nodes but also the neighbours of neighbours. For node $i$, and $\Gamma i$, which is the set of node $i$'s neighbors, we have \cite{13ye2023vital}:

\begin{equation}\label{semilocal_eq}
 SCL(i) = \Sigma_{j\in\Gamma{i}}{\Sigma_{k\in\Gamma{j}}{k}}.
\end{equation}

\textbf{Semi-local integration centrality:} It considers more features related to a node, including features of its sub-network and the weight of the degree. For each node, $i$, an edge $e(i,j)$ connects node $i$ to node $j$. Then, for each edge $e$, we count all the base cycles in which edge $e$ is involved. This measure considers the weight of edges and the degree of nodes \cite{15ban2022semi}.

\textbf{(Global) Temporal Betweenness (TB):} It is a representative measure of node importance that considers the number of times a node is located on the shortest path between two nodes. A high betweenness for a node means that more information passes through this node to reach other nodes; an attack on this node can disrupt information diffusion. The temporal betweenness finds nodes that are in the temporal shortest path between two nodes. Different algorithms are proposed for betweenness, for example, a polynomial time algorithm \cite{1temporalbetweenness-2021} and an algorithm for link streams \cite{2simard2021computing}.

\textbf{(Global) Temporal Closeness (TC):} It is the sum of the inverse of the shortest path from $i$ to all the other nodes \cite{46zhang2017degree, 47eballe2021closeness}. The Harmonic closeness algorithm is another algorithm proposed for computing the top-\textit{k} temporal closeness \cite{4_oettershagen2022computing}. Crescenzi et al. proposed an approximation for temporal closeness based on sampling and backward BFS \cite{5crescenzi2020finding}.

\subsection{Empirical networks}

We applied our methods on face-to-face interaction networks using four distinct datasets from the SocioPatterns website. These include the contact network from the SFHH conference 2019 and capturing interactions at a professional event using RFID sensors \cite{52cattuto2010dynamics}. The high school contact network from 2013 records daily student interactions through wearable RFID sensors \cite{54mastrandrea2015contact}. Additionally, we used the workplace contact network in France from the second deployment in 2015, which provides insights into employee interactions and productivity \cite{53genois2018can}. The edges in these networks show the active connections in 20s intervals. The fourth network examines contact interactions in hospital wards to understand infection transmission \cite{55vanhems2013estimating} (Table 2).

\begin{table}
\centering
\begin{tabular}{ | c | c | c | c | c | }
\hline
\bfseries Data set     	       & \bfseries $N$    &   \bfseries $T$    &\bfseries $E$ & \bfseries Duration \\\hline
High school 	   & 326 & 5818  & 188508         & 7374  \\\hline
Workplace 	       & 216 & 4274  & 78249          & 993540  \\\hline
Hospital      	       & 74   & 1139  & 32424     & 347480  \\\hline
Conference & 402  & 9565 & 70262             & 347500  \\\hline
\end{tabular}
\caption{Number of nodes ($N$), number of edges ($E$), total active time based on time steps of 20s ($T$), and the total duration, i.e. the number of time steps or snapshots.}
\end{table}\label{tb:datasetproperty}

\subsection{Epidemic models}

Epidemic models aim to reproduce an epidemic dynamic process. In the fundamental SIR epidemic network model, nodes can be in different states: $\mathcal{S}$ for Susceptible, $\mathcal{I}$ for Infected, and $\mathcal{R}$ for Recovered. When an epidemic starts, all nodes are in the $\mathcal{S}$ state except one that is infected $\mathcal{I}$; this is the seed of the epidemic. When the susceptible nodes contact the infected nodes, their state changes to $\mathcal{I}$ with probability $p$ (here, $p=1$ for simplicity), infected nodes recover with probability $q$ (here, $q=1$); once they recover, they cannot get infected again \cite{50garibaldi2020modelling}. Mathematically, being in the recovery state is thus equivalent to an effective vaccination since a node cannot get infected any longer. To estimate the diffusion speed of an epidemic outbreak, we report $di/dt$, which represents the number of newly infected nodes in each time step. We can then study the evolution of the epidemics in the network, such as the peak time and when the epidemic vanishes.

\section{Temporal centrality Measures}
\label{novel}

We introduce three novel temporal centrality measures for temporal networks. Each measure will capture different temporal-structural properties of the nodes. 

\begin{figure}
  \centering
  \includegraphics[width=.7\linewidth]{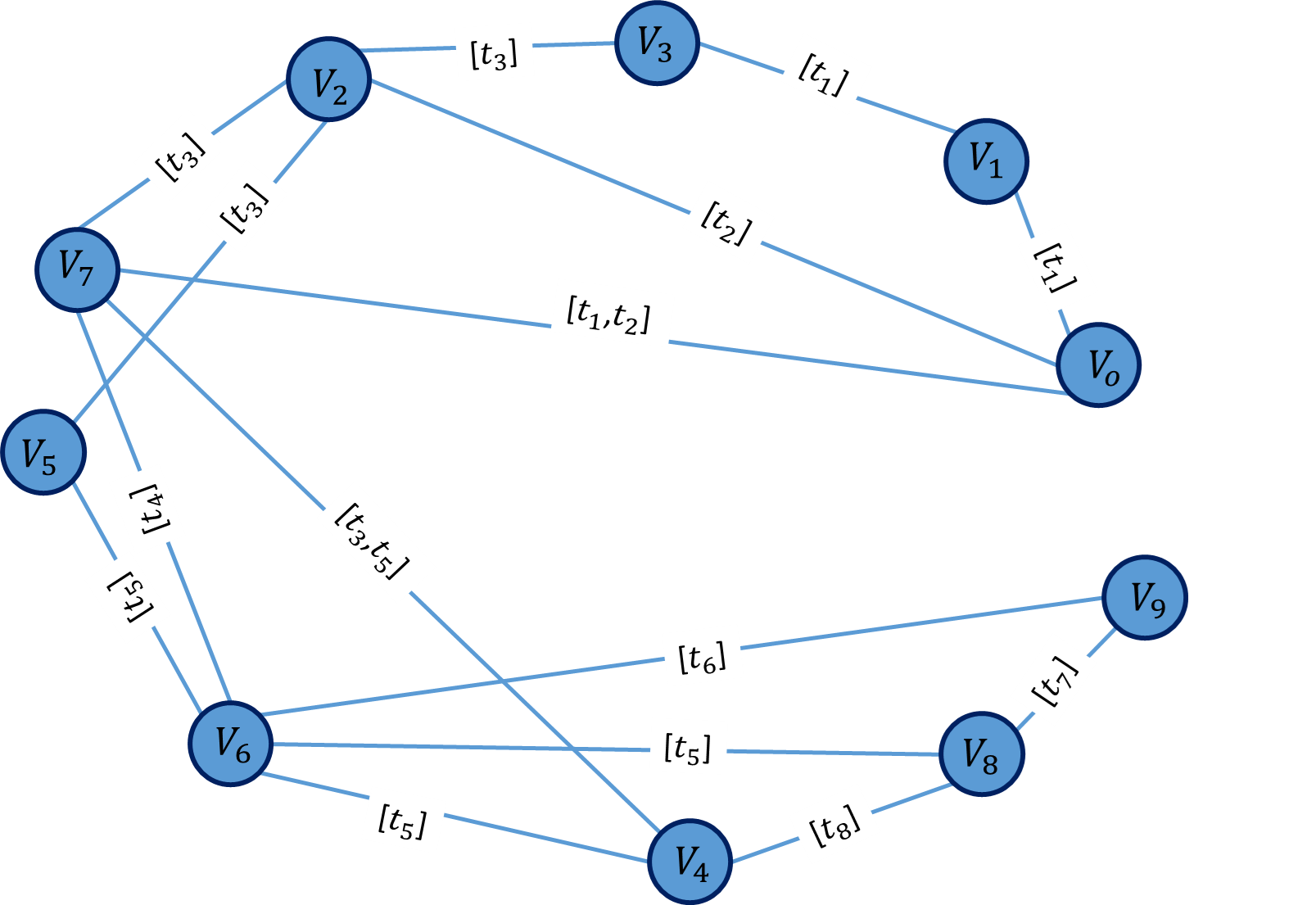}
  \caption{A simple temporal network with edge labeled with time steps.}\label{fig:graph_basesamplepng}
\end{figure}

Figure \ref{fig:graph_basesamplepng} shows a temporal network with nodes indexed from $V_0 ... V_9$. The edge labels on the network represent the time step in which the edges are connected; thus, the edges are inactive for the rest of the time. Each node in this network has a unique feature. Node $V_6$ has the highest degree, node $V_7$ has edges with the longest active time and more second-order neighbours (neighbours of neighbours) and nodes $V_2$ and $V_6$ are involved in more base cycles.

\subsection{Temporal Supra Cycle Ratio}

The temporal cycle ratio (TSCR) is a measure for detecting the most important and influential nodes based on the number of cycles in which they are involved. This measure is proposed based on the static cycle ratio \cite{13ye2023vital}. The basis of TSCR is the number of circles in which a node and its neighbours are involved.

For a node $i$:
\begin{equation}\label{tcycleRation_eq}
  TSCR(i)= \sum_{t= 0}^{T}{\sum_{j\in V}tcr_{i,j}}
\end{equation}

\begin{equation}\label{subcycleratio_eq}  tcr _{i,j}=
    \begin{cases}
      0, & \mbox{} \mathcal{AC}_{ij} = 0, \\
      \frac{\mathcal{AC}_{ij}}{\mathcal{AC}_{jj}}, & \mbox{otherwise}.
    \end{cases}
\end{equation}

where $\mathcal{AC}_{ij}$ is the number of temporal cycles in which two nodes $i$ and $j$ are involved, and $\mathcal{AC}_{jj}$ is the total number of temporal cycles of node $j$. These two parameters are the elements of the temporal cycle matrix(TCM) related to network $\mathcal{G}$.

In Figure \ref{fig:graph_basesamplepng}, if the flow starts from the node $V_0$ in the network, then at the end, the set of cycles is:
$$\mathcal{C}_{\mathcal{G}} = \biggl\{[V_7, V_ 2, V_ 0], [V_7, V_6, V_5, V_2], [V_4, V_8, V_6], [V_9, V_8, V_6], [V_7, V_4, V_6], [V_1, V_3, V_2, V_0]\biggr  \}.$$

These cycles are completed over time and are based on the sequence of nodes activated in a sequence of time steps. However, in the temporal version, the cycle set is different for each time step. For example, for time step 5, the set of cycles is:
$$\mathcal{C}_{\mathcal{G}} = \biggl\{[V_7, V_2, V_0], [V_7, V_6, V_5, V_2], [V_7, V_4, V_6], [V_1, V_3, V_2, V_0]\biggr  \}.$$

The following matrix shows the number of cycles that every two nodes involved ($\mathcal{AC}_{ij}$), and the main diagonal of the matrix is the total number of cycles that a node is involved in ($\mathcal{AC}_{jj}$).

$$
\begin{bmatrix}
2 & 1 & 2 & 1 & 0 & 0 & 0 & 1 & 0 & 0\\
1 & 1 & 1 & 1 & 0 & 0 & 0 & 0 & 0 & 0\\
2 & 1 & 3 & 1 & 0 & 1 & 1 & 2 & 0 & 0\\
1 & 1 & 1 & 1 & 0 & 0 & 0 & 0 & 0 & 0\\
0 & 0 & 0 & 0 & 1 & 0 & 1 & 1 & 0 & 0\\
0 & 0 & 1 & 0 & 1 & 1 & 2 & 2 & 0 & 0\\
0 & 0 & 1 & 0 & 1 & 1 & 2 & 3 & 0 & 0\\
0 & 0 & 0 & 0 & 0 & 0 & 0 & 0 & 0 & 0\\
0 & 0 & 0 & 0 & 0 & 0 & 0 & 0 & 0 & 0\\
\end{bmatrix}
\quad
$$

Based on eq. \ref{tcycleRation_eq}, we get $\sum_{j\in V}tcr_{i,j}$, and the final result is obtained by summing over time. Table 3 shows the values of TSCR for the sample network nodes. In TSCR, node $V_6$ is the most important node because it is involved in more cycles, followed by node $V_2$, and the rest of the nodes.

\begin{figure}[!h]
    \includegraphics[width=\textwidth]{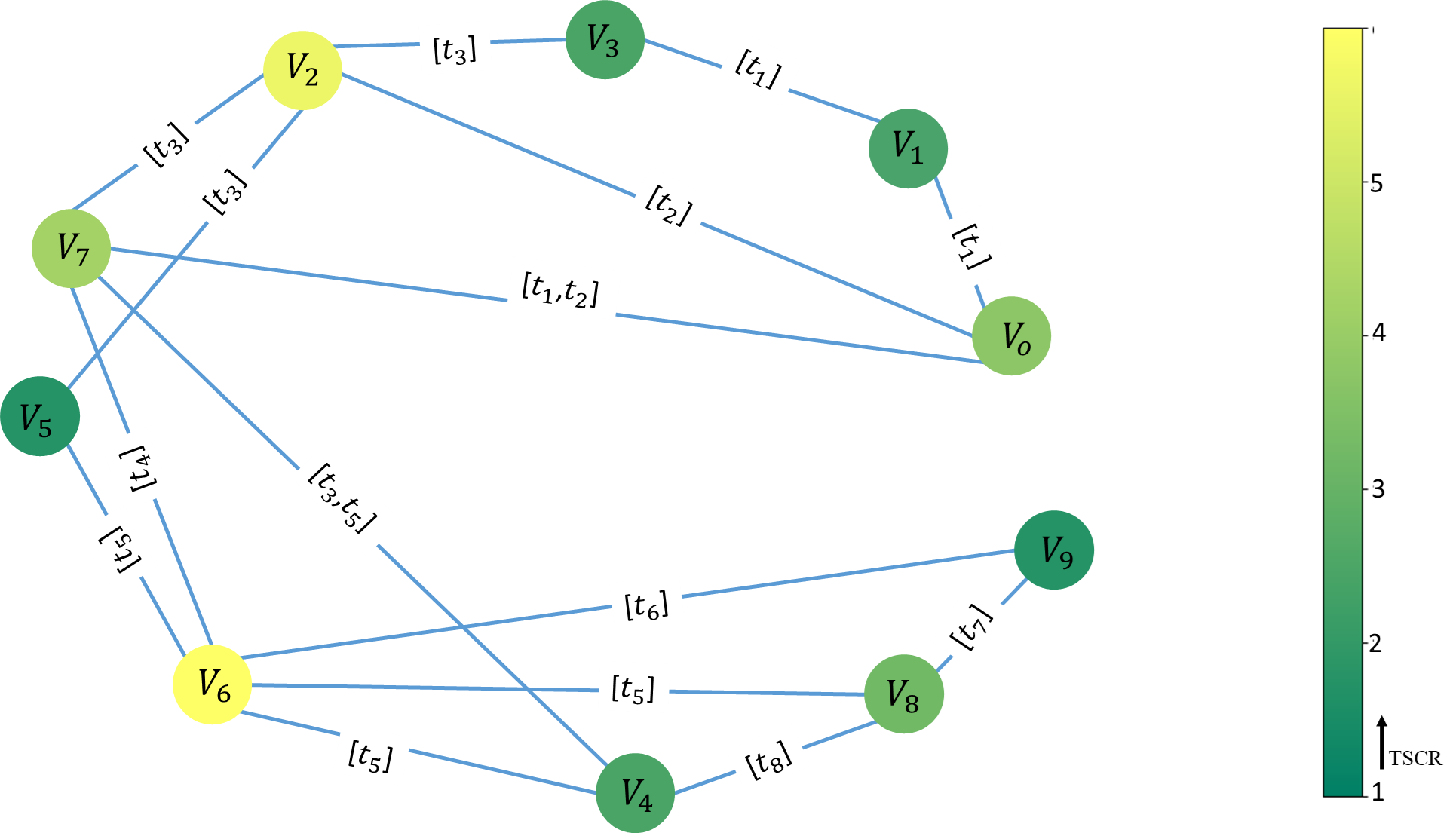}
    \centering
    \captionsetup{font=small}
    \caption{Illustration of the temporal supracycle ratio (TSCR). The most important nodes have lighter colour.}
    \label{fig:Cycleratio_plot}
\end{figure}

Figure \ref{fig:Cycleratio_plot} shows the TSCR for the simple network. The size and colour of the nodes indicate the node importance based on the TSCR index. By tracing the network cycles, we identify that nodes $V_6$ and $V_2$ are involved in more cycles; thus, nodes $V_6$ and $V_2$ are the most important nodes in the network.

\subsection{Temporal Semi-Local Integration}

Not only is the global feature of nodes important, but the local features of nodes are also important. In addition, the SLI states that a node connected to an important edge is important, and the weight of an edge shows the importance. In temporal semi-local integration (TSLI), we define the local and global features used in SLI. Therefore, we need three base differences to SLI:

\begin{enumerate}[label=(\roman*)]
        \item All nodes have the same weight.
        \item The edge degree is defined based on the total time that the connection is active.
        \item The cycle is defined based on the active connections in each time step.
    \end{enumerate}

The edge cycle factor is defined as follows:
\begin{equation}\label{eq:landa}
  \lambda(e) = P(e)+1,
\end{equation}
where $P(e)$ is the number of base cycles an edge is included. For each node $i$, TSLI is as follows:
\begin{equation}\label{eq:Ii}
  I(i) = d_{i,\mathcal{G}} + \Sigma_{j \in \Gamma_{i,\mathcal{G}}}{I(i,j)}.
\end{equation}

$I(\Gamma_{i,\mathcal{G}})$ is a set of $i$'s neighbors and $d_{i, \mathcal{G}}$ is the temporal degree of node $i$:

\begin{equation}\label{Ie}
  I(i,j) = \lambda(e). \biggl|d_{i,\mathcal{G}}+d_{j,\mathcal{G}}-2w(i,j)\biggr|.\frac{w(i,j). d_{i,\mathcal{G}}}{d_{i,\mathcal{G}}+d_{j,\mathcal{G}}}.
\end{equation}

where $w(i,j)$ is the weight of edge $(i,j)$, which is equal to the total active time of that edge. This is the best definition for edge weight because a more active edge indicates higher importance; thus, it also makes end nodes important. The following equation shows the weight of the edge $(i,j)$:
\begin{equation}\label{eq:wij}
   w(i,j) = \Sigma_{t \in T} \mathcal{L}(i,j,t),
\end{equation}
where $T$ is the total time window, we trace the network behaviour.

This measure represents the integrity of each node in the neighbourhood. As long as a node is involved in more cycles, it has a denser neighbourhood.

Since we consider the weight of all nodes equal to one, it is possible that $ (d_{i,\mathcal{G}}+d_{j,\mathcal{G}}-2w(i,j))$ in equation \ref{Ie} gives a negative result, and as much it is involved in different cycles, it becomes more negative and less important. Therefore, we use the absolute value of $ (d_{i,\mathcal{G}}+d_{j,\mathcal{G}}-2w(i,j))$ in equation \ref{Ie}.

Figure 3 shows the TSLI value for the sample network. Here, node $V_7$ is the most important, followed by node $V_6$. Node $V_7$ is connected to two edges with a high active time; therefore, based on the idea that the node connected to the important edge is also important, the TSLI value of $V_7$ is higher than that of the other nodes. Node $V_6$ is also important because it is involved in more cycles.

\begin{figure}[h]
    \includegraphics[width=\textwidth]{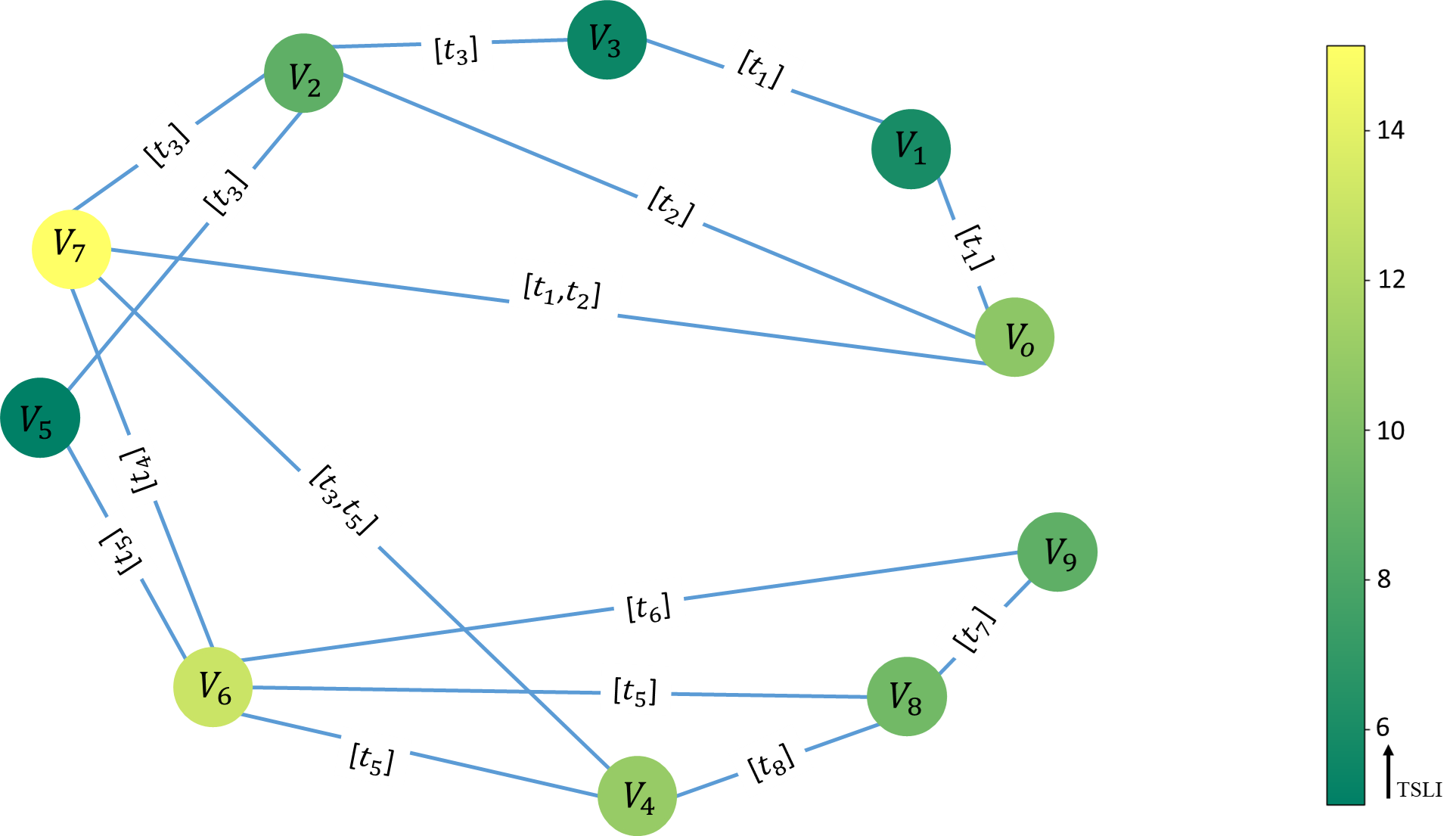}
    \centering
    \captionsetup{font=small}
    \caption{Illustration of important nodes using Temporal Semi-Local Integration (TSLI). The most important nodes have lighter colour.}
    \label{fig:integration_plot}
\end{figure}

Table \ref{tbl:samplenet} shows the TSLI (eq.~\ref{eq:Ii}) for all nodes. A high TSLI indicates that a node is critical in the network. Compared with Figure \ref{fig:integration_plot}, the nodes connected to the more important nodes also have higher TSLI values. 

\subsection{Temporal Semi-Local Centrality}

Semi-local centrality is based on the neighbours and neighbours of neighbours for node $j$. Therefore, in static networks, the semi-local centrality counts the second-order neighbours of a node. The TSLC is the SLC measure in the temporal network. Because the connections in the temporal networks are temporal, the TSLC for node $j$ is all the second-order neighbours that are reachable according to consecutive time steps starting at time $t$. For node $j$, we have:

\begin{equation}\label{temporal_sei_local_eq}
  TSLC_{t,\mathcal{G}}(j) = \{i :  \exists i' | (i,i',t) \in \mathcal{E} \&  \exists t'>t | (i',j,t') \in \mathcal{E} \}.
\end{equation}

The total TSLC(j) for all snapshots is:

\begin{equation}\label{temporal_sei_local2_eq}
 TSLC_{\mathcal{G}}(k) = \sum_{t=0}^{T}{\sum_{k\in {\Gamma_{i,\mathcal{G}(t,T)}}}{\sum_{z\in{\Gamma_{k,\mathcal{G}(t+1,T)}} }{z}}}.
\end{equation}

We call this measure the semi-local centrality measure because it considers the importance of a node in a wider area than local. The degree centrality index is strictly based on neighbours, but the semi-local centrality extends to neighbours of the neighbours.

Table \ref{tbl:samplenet} shows the TSLC for the nodes in the sample network. This figure shows nodes $V_7$ and $V_6$ as the most important. Because this measure is based on the second-order neighbours of the nodes, we expect that these nodes will have more second-order neighbours.

\begin{table}
\centering
\begin{tabular}{ | c | c | c | c | c | c | c | c | c | c | c | }
\hline
\bfseries Node & $V_0$ & \bfseries$V_1$ & \bfseries$V_2$ & \bfseries$V_3$ & \bfseries$V_4$ & \bfseries$V_5$ & \bfseries$V_6$ & \bfseries$V_7$ & \bfseries$V_8$ & \bfseries$V_9$\\\hline
 \bfseries TSCR & 3.88 & 2.53 & 5.71 & 2.53 & 2.46  & 1.82 & 6.11 & 4.22 & 3.24 & 1.82  \\\hline
\bfseries TSLI & 7.99 & 3.33 & 5.86 & 2.66 & 8.79  & 2 & 10.2 & 12.13 & 7 & 6 \\\hline
\bfseries TSLC & 4 & 2 & 4 & 2 & 4 & 2 & 5 & 6 & 3 & 2  \\\hline
 \end{tabular}
\caption{TSCR value for all nodes in the network of Figure \ref{fig:graph_basesamplepng}}\label{tbl:samplenet}
\end{table}

Figure 4 represents the node importance based on TSLC. Similar to the TSLI, nodes $V_7$ and $V_6$ are the most important nodes since they have more second-order neighbours, node $V_7$ with nine second-order neighbours including \{$V_0$,$V_1$,$V_2$,$V_5$,$V_6$,$V_8$\} and node $V_6$ with six second-order neighbours \{$V_0$,$V_2$,$V_4$,$V_7$,$V_8$,$V_9$\}. Both nodes have higher degrees, and their connections are more active than others, which are also in contact with nodes with high degrees.
 
\begin{figure}[!h]
    \includegraphics[width=\textwidth]{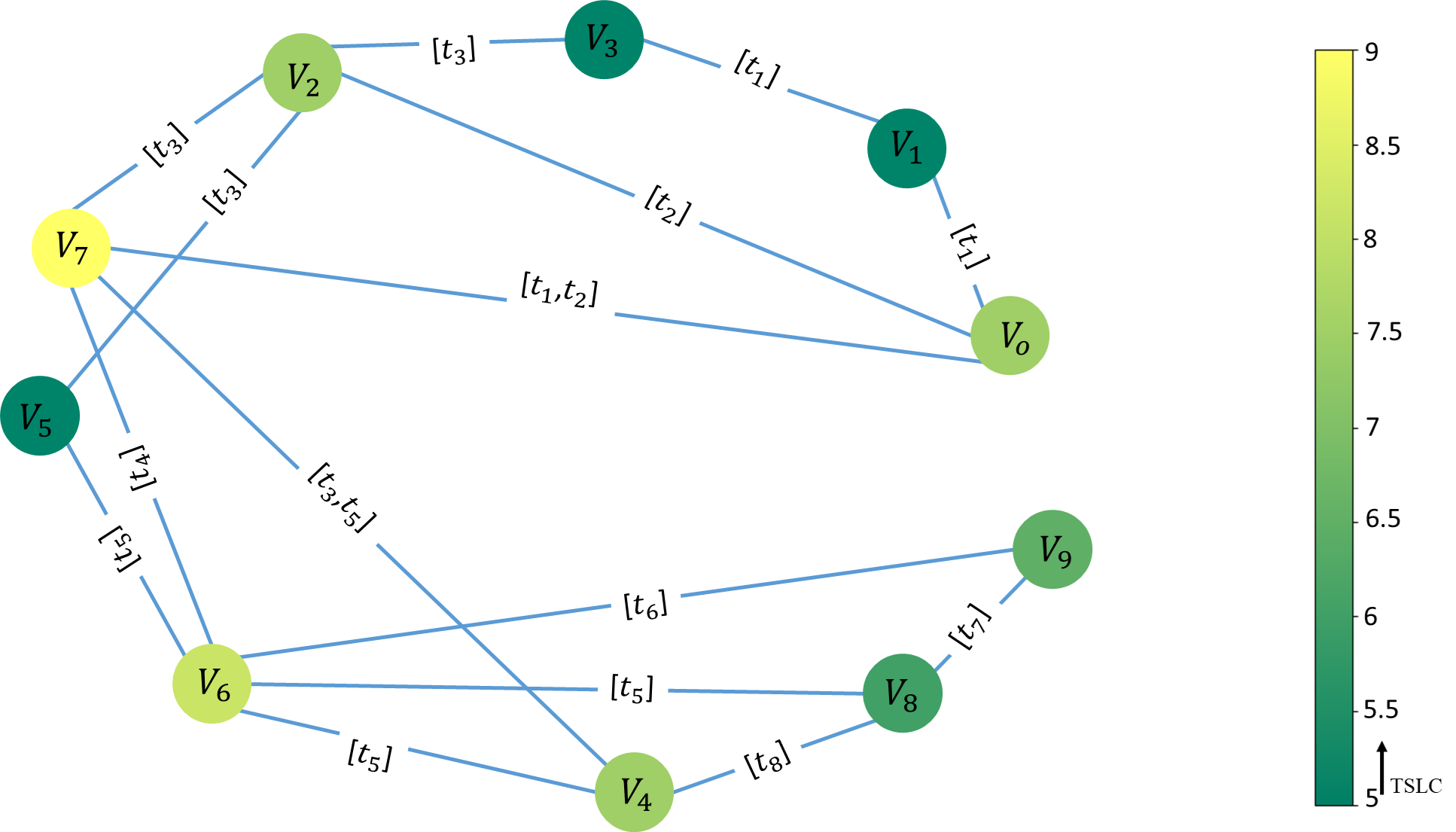}
    \centering
    \captionsetup{font=small}
    \caption{Illustration of important nodes using Temporal Semi-Local Centrality (TSLC). The most important nodes have lighter colour.}
    \label{fig:centrality_plot}
\end{figure}

\section{Results}
\label{results}

We first check and compare the detected nodes by all six measures. The comparison includes three cases: the SIR propagation speed when the critical nodes are spreaders, the SIR propagation speed when the critical nodes are removed, and the largest connected component.

\subsection{Epidemic Spread}

When discussing epidemic spread, we face two problems: controlling and predicting the epidemic. When an epidemic outbreak occurs, we must isolate critical nodes, i.e. the nodes that regulate the epidemic spread, to prevent the epidemics. Both these approaches prompted us to analyse the effect of critical nodes on epidemic spread speed.

\begin{figure}[H]
    \centering
    \begin{tabular}{cccc}
        \begin{subfigure}[b]{0.22\textwidth}
            \centering
            \includegraphics[width=\textwidth]{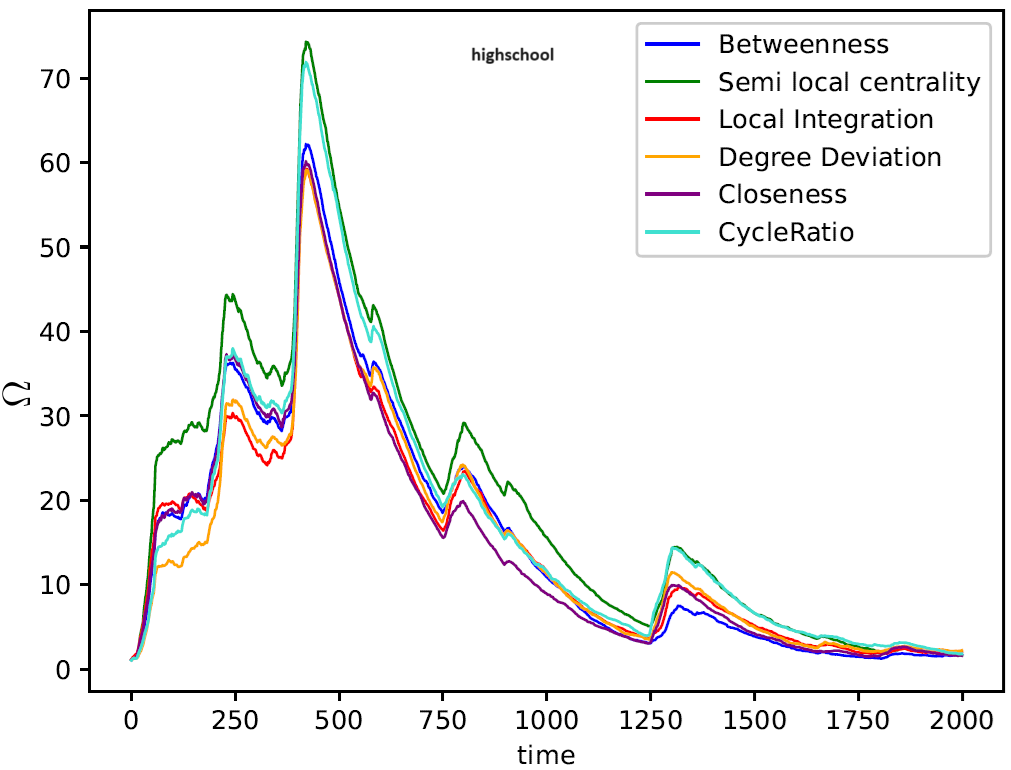}
            \caption{}
            \label{fig:plot1}
        \end{subfigure}\hspace{-4mm} &
        \begin{subfigure}[b]{0.22\textwidth}
            \centering
            \includegraphics[width=\textwidth]{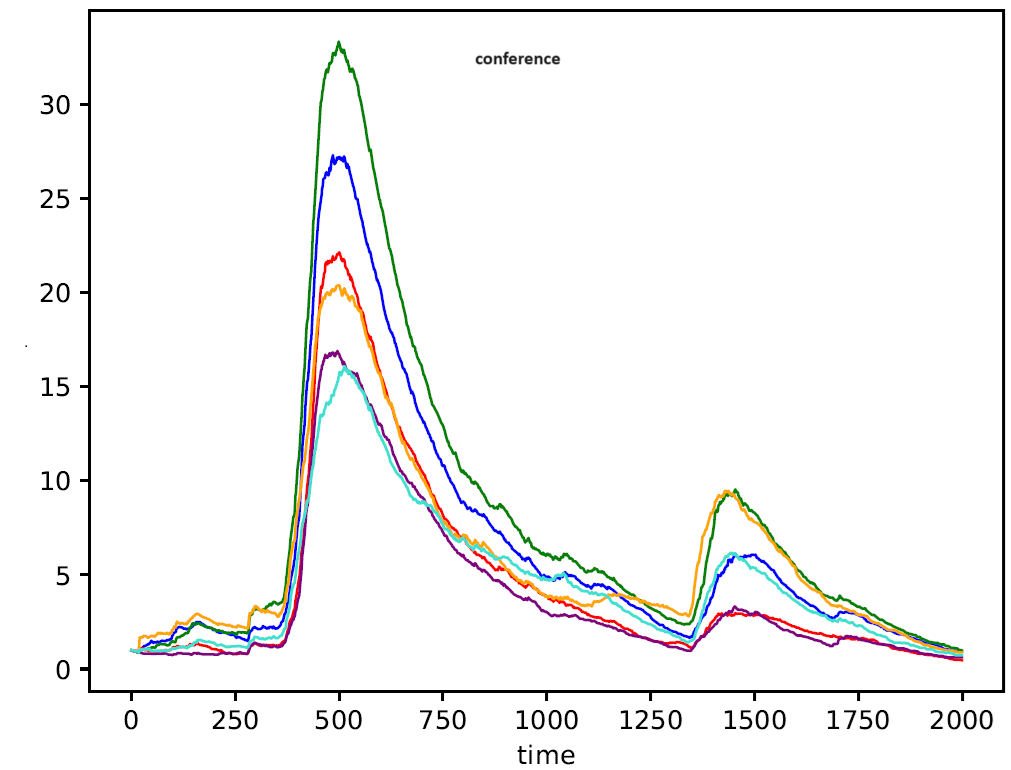}
            \caption{}
            \label{fig:plot2}
        \end{subfigure}\hspace{-4mm} &
        \begin{subfigure}[b]{0.22\textwidth}
            \centering
            \includegraphics[width=\textwidth]{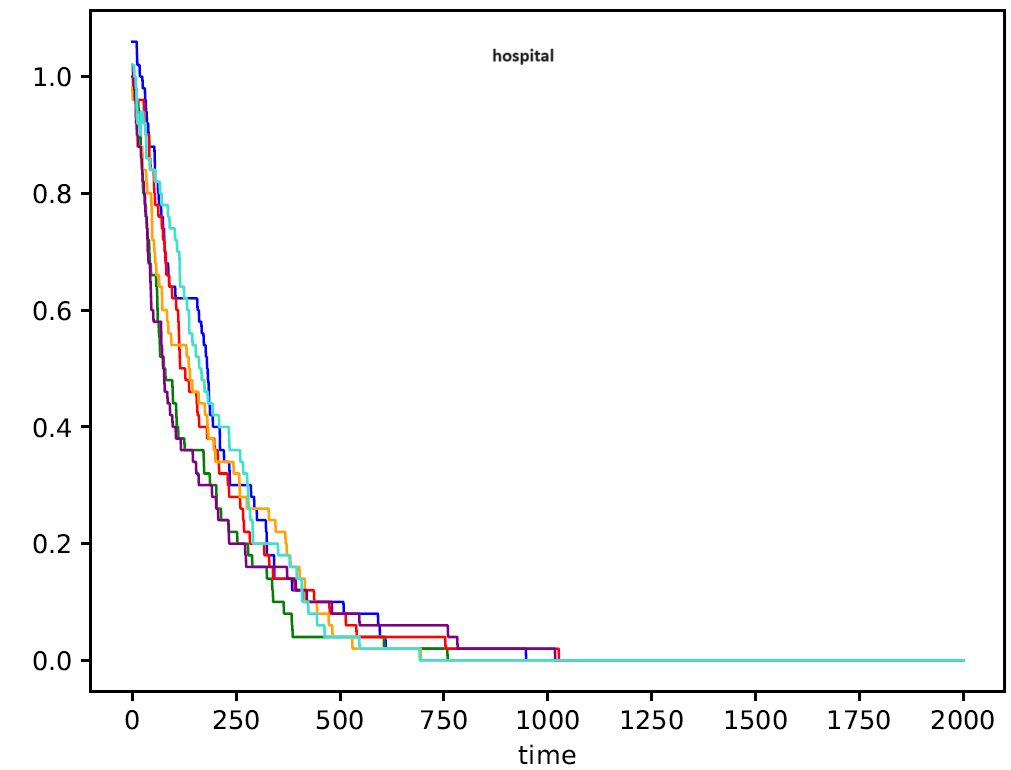}
            \caption{}
            \label{fig:plot3}
        \end{subfigure}\hspace{-4mm} &
        \begin{subfigure}[b]{0.22\textwidth}
            \centering
            \includegraphics[width=\textwidth]{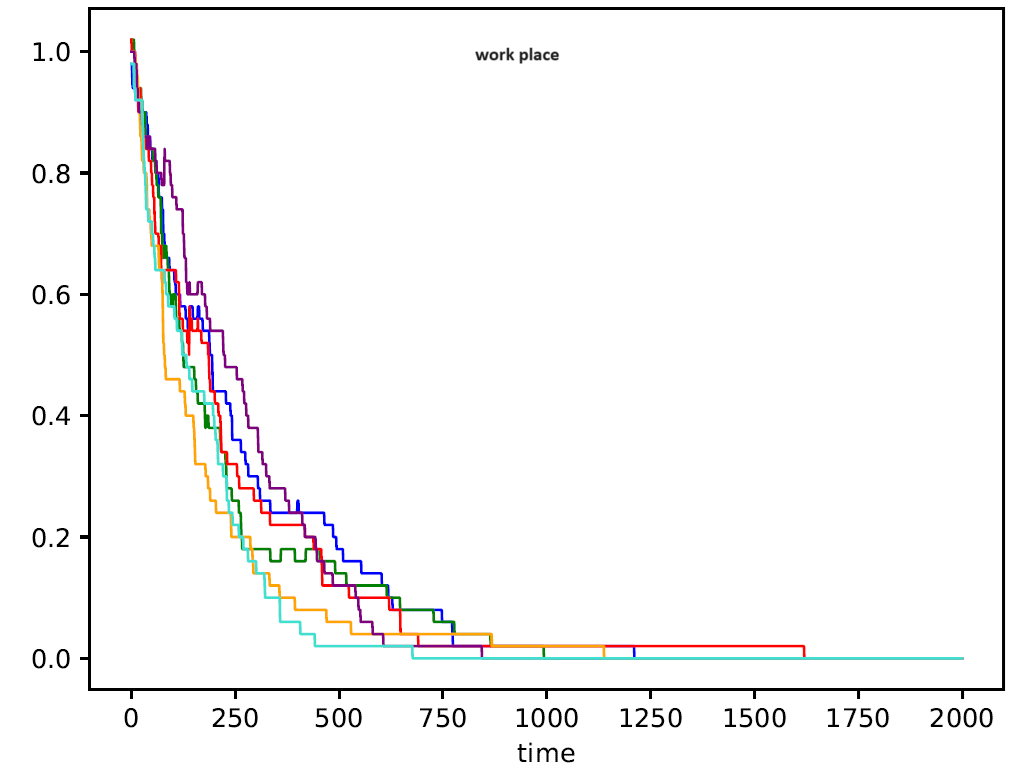}
            \caption{}
            \label{fig:plot4}
        \end{subfigure} \\
        \begin{subfigure}[b]{0.22\textwidth}
            \centering
            \includegraphics[width=\textwidth]{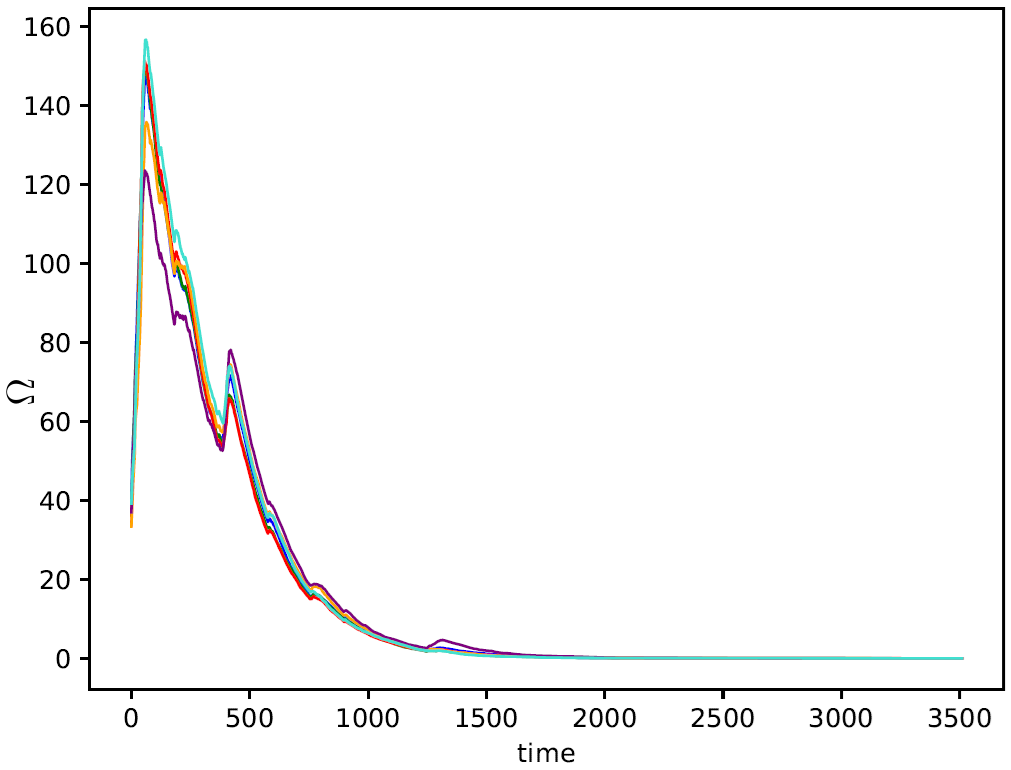}
            \caption{}
            \label{fig:plot5}
        \end{subfigure}\hspace{-4mm} &
        \begin{subfigure}[b]{0.22\textwidth}
            \centering
            \includegraphics[width=\textwidth]{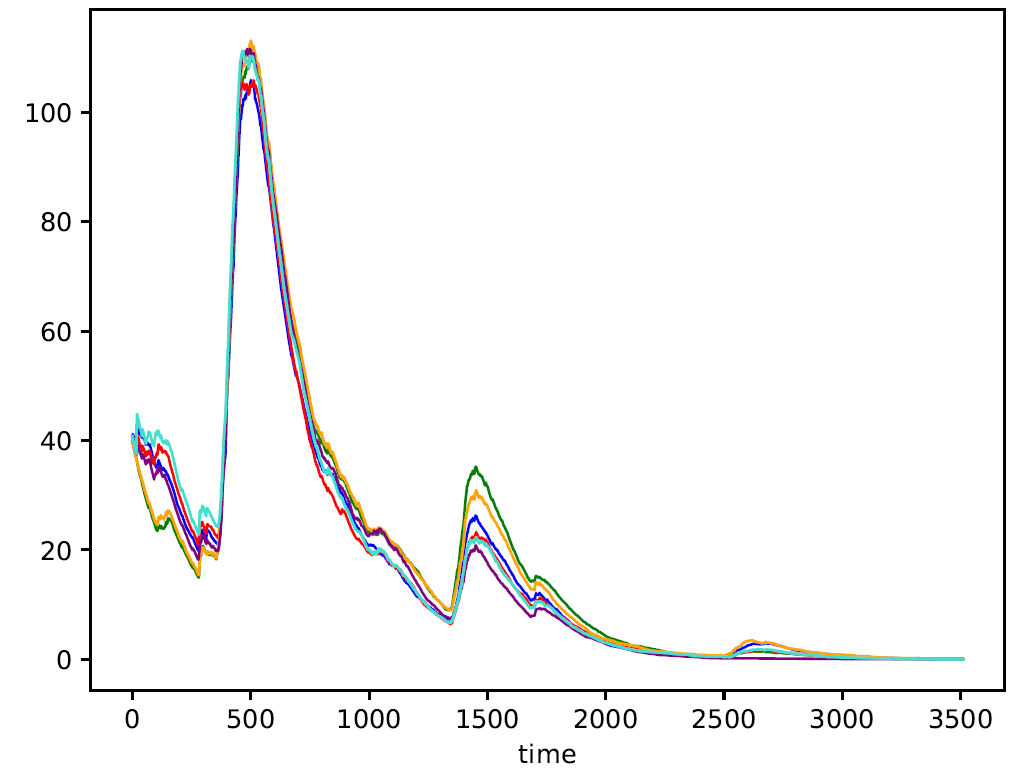}
            \caption{}
            \label{fig:plot6}
        \end{subfigure}\hspace{-4mm} &
        \begin{subfigure}[b]{0.22\textwidth}
            \centering
            \includegraphics[width=\textwidth]{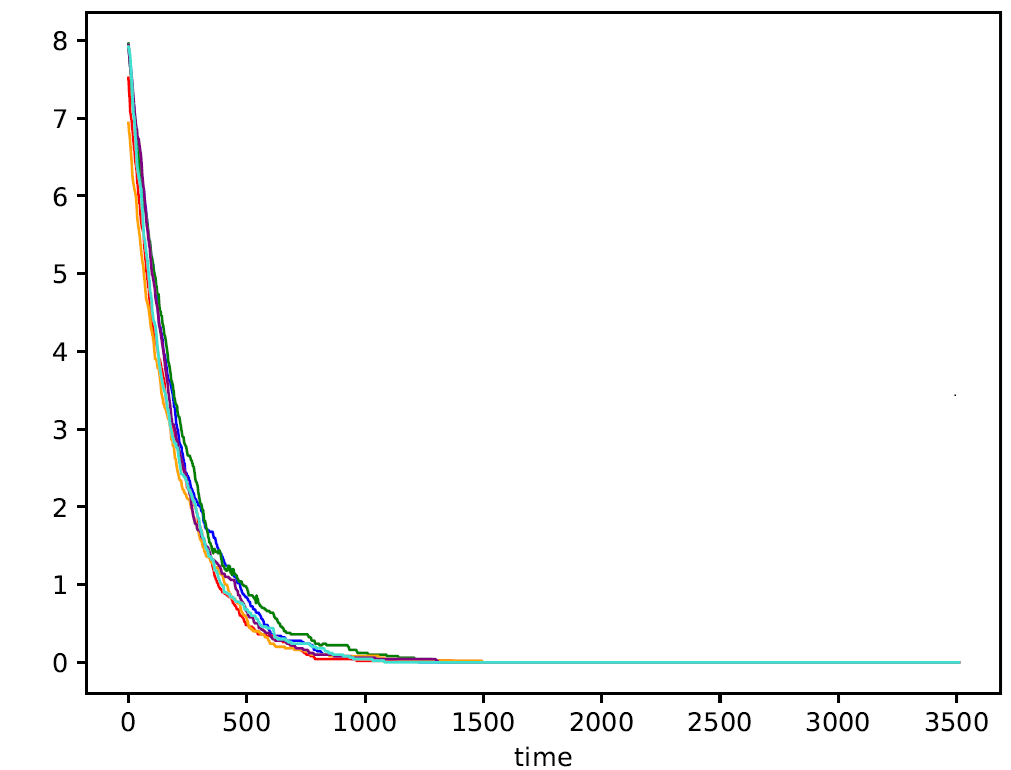}
            \caption{}
            \label{fig:plot7}
        \end{subfigure}\hspace{-4mm} &
        \begin{subfigure}[b]{0.22\textwidth}
            \centering
            \includegraphics[width=\textwidth]{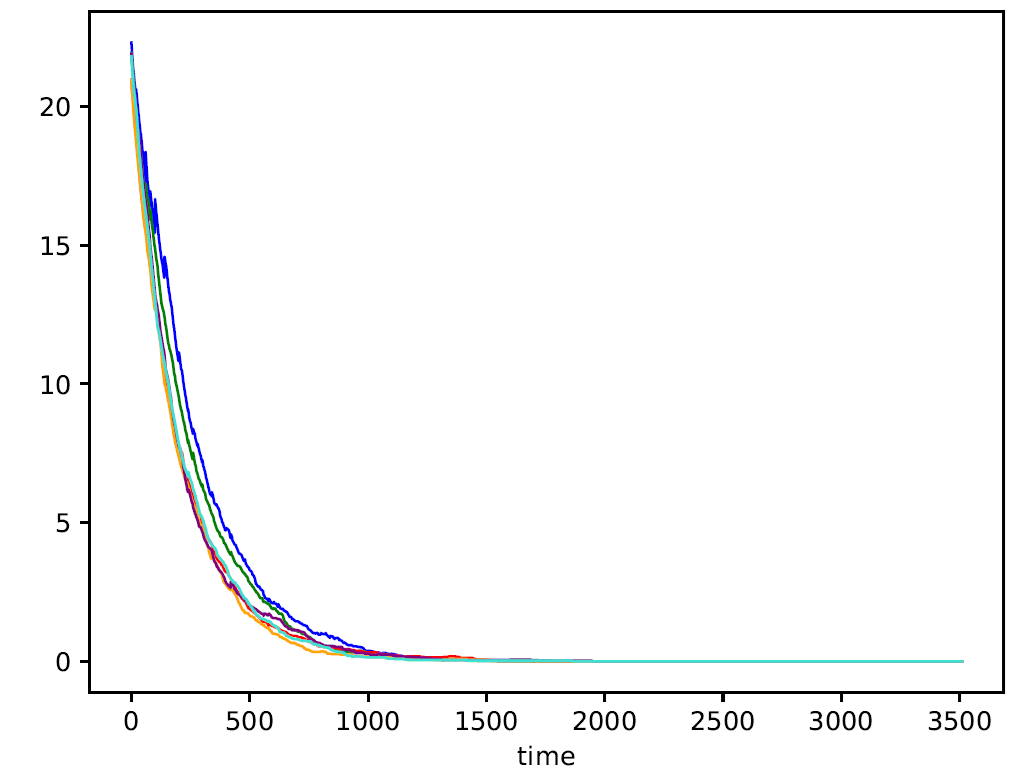}
            \caption{}
            \label{fig:plot8}
        \end{subfigure}
    \end{tabular}
    \caption{(a-d) Epidemic spread after removing the top $1\%$ critical nodes for isolation and then comparing the six measures in different networks. (e-h) Epidemic spread in the network when the top $1\%$ nodes are selected as initial spreaders or seeds in the different networks.}
    \label{fig:sir_remove_and_speed}
\end{figure}

To analyse the effect of the critical nodes in the epidemic's spreading, we can isolate them or consider them the seed nodes. We use both approaches to evaluate the proposed measures' performance and compare them with known measures. At first, we remove critical nodes detected by each measure, then run the $SIR$ epidemic model in the network. For this simulation, the initially infected nodes are randomly chosen, and 
to mitigate the seed's choice effect, we repeated the simulation 50 times and reported the $\Omega = 1/M \times \sum_x di/dt$ where $M$ is the number of runs. Figure \ref{fig:sir_remove_and_speed}(a-d) shows the epidemic spread for the four datasets. 

The first peak is critical to help control the epidemic spread. Figure
\ref{fig:sir_remove_and_speed}(a-d) shows that the cycle ratio has the best performance since it shows the lowest value for $\Omega$. Regarding the hospital dataset, the first peak occurs at the beginning, and all the measures behave similarly, but the semi-local centrality decreases faster than other measures. Based on the semi-local centrality and cycle ratio, the epidemic ends sooner. In the conference dataset, the three proposed measures have the best functionality for the first peak. For the highest peak, the value of $\Omega$ for the cycle ratio is the lowest, indicating that it has the best recognition for the critical nodes. For the high school dataset, the cycle ratio and degree deviation have the lowest value for the first peak. The local integration has the lowest value in all phases (first peak, highest peak, and ending of the epidemic). Finally, in the workplace dataset, similar to the hospital dataset, all the measures for the first and highest peak behave similarly. Still, the cycle ratio decreases faster and ends the epidemic sooner than the other measures.

The other viewpoint is using critical nodes to increase the epidemic speed. In this case, we chose the most critical nodes as the initial spreaders instead of choosing randomly and then ran the $SIR$ epidemic model. In Figure \ref{fig:sir_remove_and_speed}(e-h), the initial spreaders are set to be the top $1\%$ of critical nodes detected by each of the six measures. At each time step, the epidemic spreads from infected nodes to healthy nodes that are in contact with them during that time, with a probability of $1$. The epidemic continues until all individuals in the community are infected and no new nodes are left to infect, at which point $\Omega$ becomes zero. 

The role of initial spreaders is seen in the timing of the epidemic peak and when the epidemic becomes widespread. The sooner and higher the peak occurs, the more critical the initial spreaders were in propagating the epidemic, leading to the population getting infected sooner. In the high school dataset, closeness shows the lowest value for $\Omega$. In the conference dataset, while betweenness has a low value at the beginning, it has a high value at the highest peak, indicating that the nodes infected in the second stage are more critical. In data sets related to the hospital and workplace, the epidemic had the highest value initially, but with TSCR and TSLI, the epidemic reached the whole network sooner. In the workplace data set, TDD and TSCR have the best performance, reaching the entire network. In all cases, TSCR, TSLI, and TSLC achieved the best performance in reaching the peak and infecting the whole network.  

In the next experiment, different fractions of critical nodes, ranging from $0.1$ to $0.9$, are removed. In this simulation, the seed nodes are chosen randomly, and the reported value is the average of $M=50$ runs. Table \ref{tbl:SIR_remove_percent_speed} reports the peak value of $\Omega$.

Removing critical nodes decreases the epidemic speed since they are essential in regulating the epidemic. A lower $\Omega$ indicates that the removed nodes were more critical. In most cases, the minimum values are observed for TSCR, TSLC, and TSLI, while the maximum values are mostly for TB and TDD. The reported peak values for different measures are close to the workplace dataset. Nonetheless, TSLI performs better, as it has the minimum value in five cases.

\begin{table}
\textsmaller{\textsmaller{\textsmaller{\textsmaller{\textsmaller{
\begin{center}
\renewcommand{\arraystretch}{1.5}
\setlength{\tabcolsep}{2pt}
\begin{tabular}{cccccccccccccc}

\hline

&\multicolumn{6}{c}{\textbf{high-school}}&&\multicolumn{6}{c}{\textbf{conference}}\\

\cline{2-7}

\cline{9-14}

N&TB&TSLC&TDD&TSLI&TC&TSCR&& TB&TSLC&TDD&TSLI&TC&TSCR\\\hline

0.1 &1763.72&1573.98&1701.64&1421.4&\textbf{1309.16}&1443.72&&\textbf{171.18} & 344.36 & 511.42 & 307.86 & 373.7 & 831.76\\

0.2 &4888.16&4079.52&5115.46&4567.06&4232.56&\textbf{4049.2}&& 1419.78 & 1219.24 & 1580.56 & \textbf{1013.72} & 1107.36 & 1473.36\\

0.3 &8312.82&6777.66&\textbf{5849.78}&7072.6&6744.26&6727.34&& 2246.84 & 2212.98 & 2648.32 & \textbf{1707.94} & 2488.7 & 1939.36\\

0.4 &11432.6&9841.62&11392.78&9820.94&9580.22&\textbf{9363.26}&& \textbf{2391.7} & 2727.4 & 3488.9 & 2397.1 & 4600.36 & 2688.76\\

0.5&14910.1&12605.04&14469.24&\textbf{12181.58}&12330.9&12338.74&& \textbf{3238.32} & 3287.9 & 4657.52 & 3359.54 & 5656.44 & 3442.06\\

0.6&18284.22&14887.46&16832.04&\textbf{14409.92}&15001.18&14935.96&& \textbf{3871.1} & 4303.78 & 5953.64 & 4226.1 & 6912.64 & 4249.54\\

0.7&21241.94&17281.86&19343.34&\textbf{16975.7}&17059.06&17128.78&& 4953.78 & 5197.4 & 6820.94 &\textbf{4931.68} & 8252.52 & 5094.18\\

0.8&23892.26&19527.88&22148.88&19366.62&19029.02&\textbf{18955.88}&& 5735.62 & 5570.08 & 7415.48 & \textbf{5164.1} & 9046.52 & 5665.48\\

0.9&26835.46&22026.92&24736.18&21554.12&\textbf{20625.58}&20632.20&& 6337.06 & 5947.0 & 7805.3 & \textbf{5561.74} & 10341.96 & 6208.9\\\hline


&\multicolumn{6}{c}{\textbf{hospital}}&&\multicolumn{6}{c}{\textbf{workplace}}\\

\cline{2-7}

\cline{9-14}

N&TB&TSLC&TDD&TSLI&TC&TSCR&& TB&TSLC&TDD&TSLI&TC&TSCR\\\hline

0.1 &26.18 &\textbf{ 25.5} & 27.86 & 25.84 & 26.06 & 25.62&& 24.64 & \textbf{25.1} & 25.66 & 25.5 & 26.42 & 25.5\\

0.2 &77.32 & \textbf{75.5} & 78.92 & 76.5 & 75.62 & 76.74&& 75.3 & \textbf{73.98} & 75.5 & 74.5 & 77.46 & 75.5\\

0.3 &127.6 & \textbf{123.5} & 129.5 & 126.5 & 125.38 & 127.5&& 125.5 & \textbf{122.5} & 125.94 & 124.06 & 126.5 & 125.5\\

0.4 &179.24 & \textbf{175.46} & 179.5 & 176.5 & 175.5 & 177.02&& 175.5 & \textbf{172.5} & 175.2 & \textbf{172.5} & 176.5 & 175.02\\

0.5&229.36 & 224.5 & 229.42 & 226.5 & \textbf{223.5} & 226.5&& 225.5 & 223.08 & 223.3 & \textbf{221.96} & 226.5 & 223.56\\

0.6&279.5 & 274.5 & 277.64 & 276.5 & \textbf{273.5} & 276.5&& 275.5 & \textbf{272.6} & 272.92 & 271.5 & 276.5 & 273.5\\

0.7&329.5 & \textbf{324.5} & 327.5 & 326.5 & 324.62 & 327.62 && 324.88 & 323.5 & 323.4 & \textbf{321.04} & 326.5 & 324.5\\

0.8&380.56 & \textbf{374.5} & 377.5 & 376.5 & \textbf{374.5} & 380.32&& 374.5 & 373.5 & 372.5 & \textbf{369.5} & 376.5 & 374.96\\

0.9&431.64 & \textbf{424.5} & 427.5 & 426.5 & 424.78 & 429.78&& 424.06 & 423.72 & 422.5 & \textbf{419.5} & 426.5 & 425.78\\\hline

\end{tabular}
\caption{Peak value of $\Omega$ after removing the top $x \%$ of critical nodes for isolation, and then comparing the six measures ($x\in\{0.1,0.2,...,0.9\}$).}\label{tbl:SIR_remove_percent_speed}
\end{center}
}}}}}
\end{table}

Some of the measures have more accurate detection depending on the network. However, well-known measures like betweenness, closeness, and degree deviation focus on only one of the node's features. In contrast, TSLC, TSLI, and TSCR consider a combination of node features. In the workplace and high school networks, TSLC and TSCR detect the most influential nodes because, in these two networks, the nodes have the most influence on their semi-local neighbours within their communities. In these networks, it is rare for a node to have a global effect; usually, it affects a group of friends. Therefore, we expect the detected nodes to have less influence on betweenness and closeness, which consider the global features of nodes. On the other hand, at conferences where people try to connect with others and form new relationships, measures like betweenness and closeness, which consider the global features of nodes, have the best functionality. In contrast, TSCR, which focuses on semi-local features, is less valuable. Finally, in the hospital dataset, where connections are more uniform, all the measures exhibit similar performance.

\subsection{Network Robustness}

In this section, we study the impact of node removal on network fragmentation and evaluate network robustness through percolation theory. As the network becomes denser, the removal of nodes tends to have less effect on the size of the largest connected component($S_{max}$), indicating higher network robustness [69]. Thus, assessing the critical node detection through percolation theory provides valuable insights into network resilience \cite{44li2021percolation, 45badie2022directed,13ye2023vital}.

\begin{figure}[H]
    \centering
    \begin{tabular}{cccc}
        \begin{subfigure}[b]{0.22\textwidth}
            \centering
            \includegraphics[width=\textwidth]{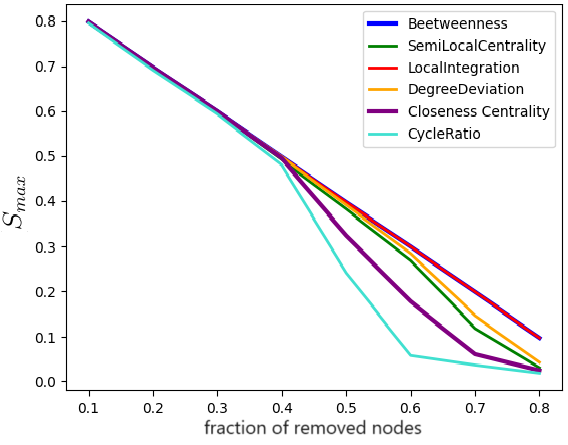}
            \caption{}
        \end{subfigure}\hspace{-4mm} &
        \begin{subfigure}[b]{0.22\textwidth}
            \centering
            \includegraphics[width=\textwidth]{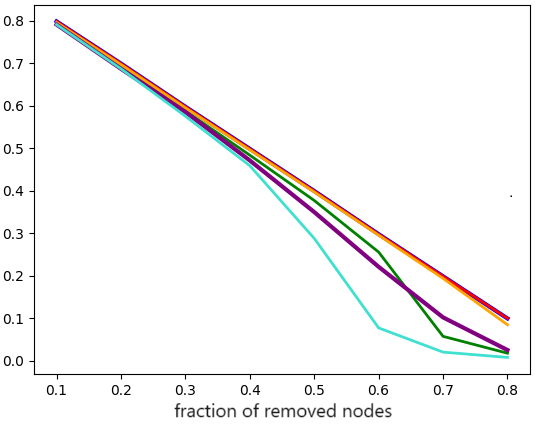}
            \caption{}
        \end{subfigure}\hspace{-4mm} &
        \begin{subfigure}[b]{0.22\textwidth}
            \centering
            \includegraphics[width=\textwidth]{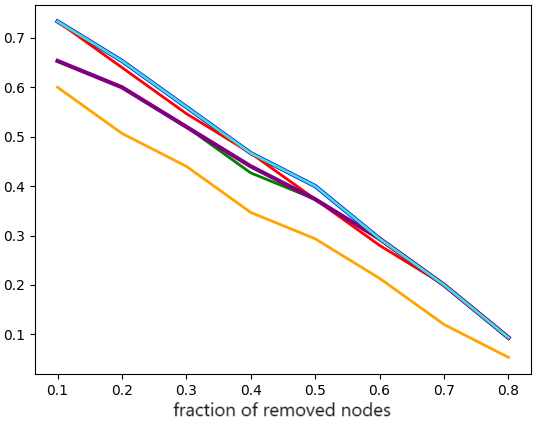}
            \caption{}
        \end{subfigure}\hspace{-4mm} &
        \begin{subfigure}[b]{0.22\textwidth}
            \centering
            \includegraphics[width=\textwidth]{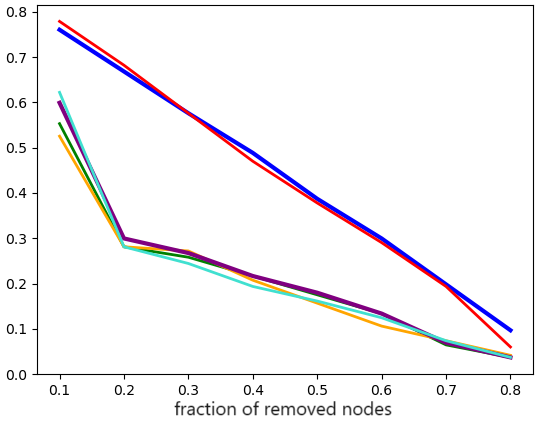}
            \caption{}
        \end{subfigure} \\
        
    \end{tabular}
    \caption{Comparison of $S_{max}$ after removing critical nodes based on six measures in different dataset: (a) high-school, (b) conference, (c) hospital, (d) workplace.}
    \label{fig:gcc}
\end{figure}

To compare the accuracy of critical node detection, we order nodes based on six measures of importance. Subsequently, we iteratively remove nodes according to their importance, reporting the largest connected component size for all four networks (Fig.~\ref{fig:gcc}). Across all datasets, temporal closeness centrality and temporal supra-cycle ratio exhibit similar behavior, showing superior performance in the high-school and conference datasets by inducing more disconnections, leading to smaller sizes of the largest connected component. In the workplace and hospital datasets, degree deviation demonstrates the best performance, while it performs moderately in the other two datasets. Temporal supra-cycle ratio and temporal semi-local centrality exhibit the best performance overall, while temporal local integration centrality and temporal betweenness centrality show similar behaviour. The impact of removing critical nodes differs significantly from that of marginal nodes. Removing several nodes with less importance yields a different effect than removing a critical node; even removing a small fraction of essential nodes can result in network disconnection. Therefore, the effectiveness of a measure lies in its ability to identify critical nodes accurately. In our experiment, the supra-cycle ratio demonstrates the most reliable performance across all four networks, while other measures show promising results in different networks.

\subsection{Correlation Analysis}

\begin{figure}[htbp]
  \centering
  \begin{subfigure}[b]{\textwidth}
    \centering
    \begin{subfigure}[b]{0.45\textwidth}
      \includegraphics[width=\textwidth]{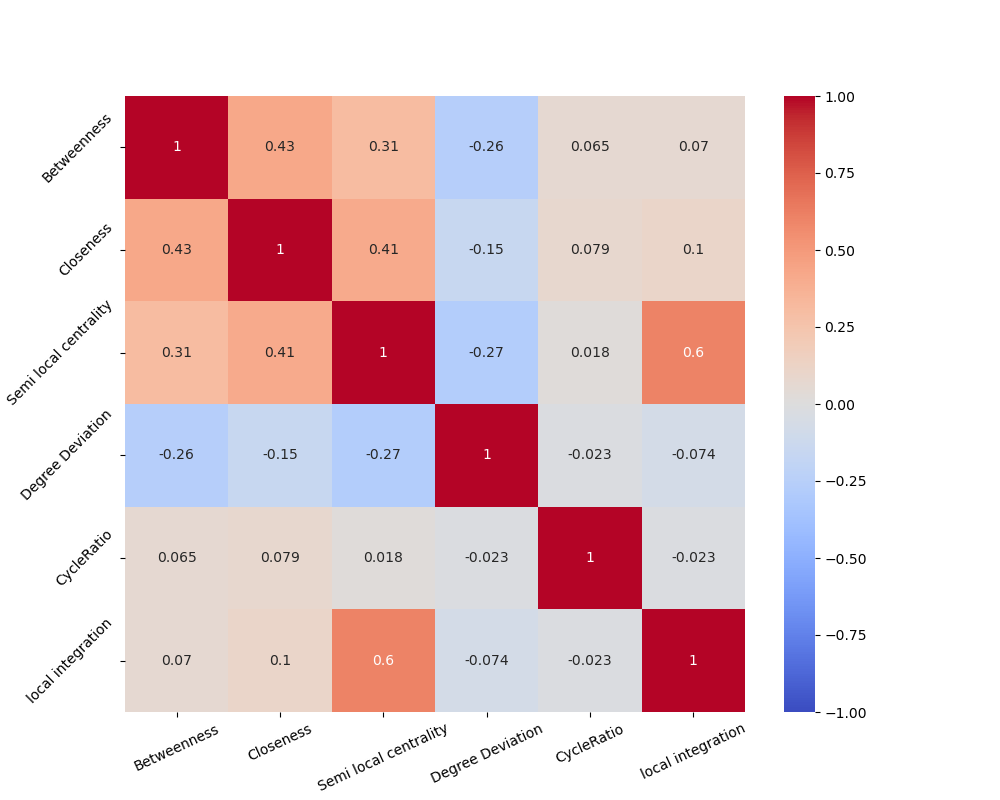}
      \caption{}
      \label{fig:subfig1}
    \end{subfigure}
    \begin{subfigure}[b]{0.45\textwidth}
      \includegraphics[width=\textwidth]{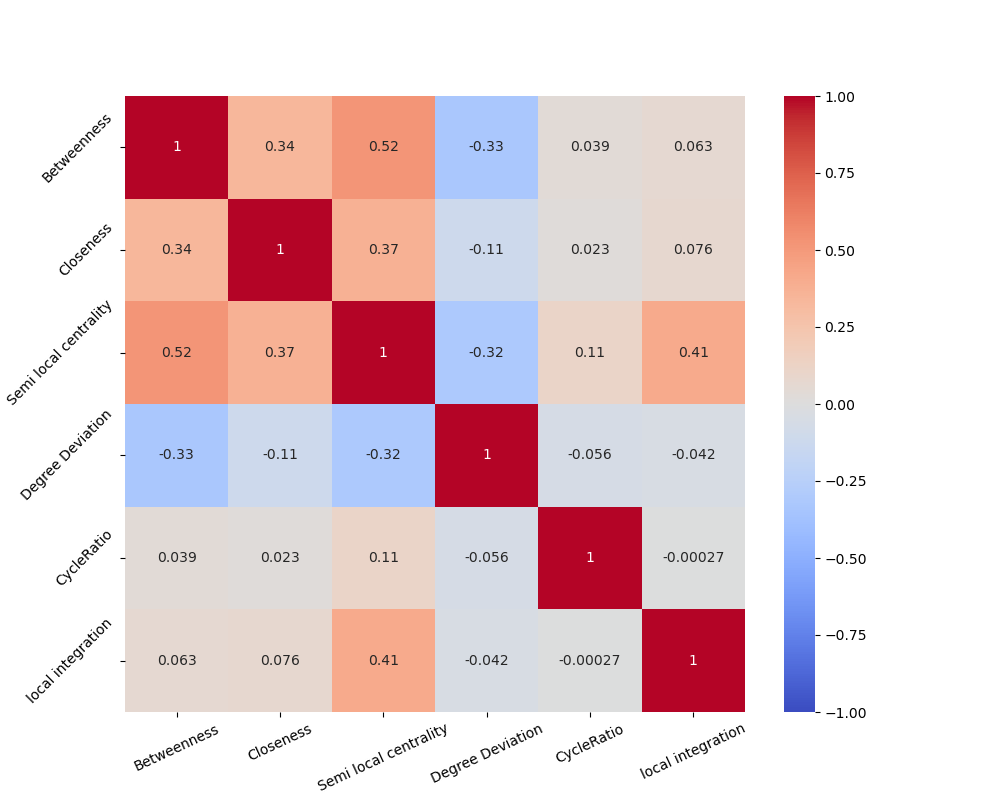}
      \caption{}
      \label{fig:subfig2}
    \end{subfigure}
    \label{fig:pair1}
  \end{subfigure}

  \begin{subfigure}[b]{\textwidth}
    \centering
    \begin{subfigure}[b]{0.45\textwidth}
      \includegraphics[width=\textwidth]{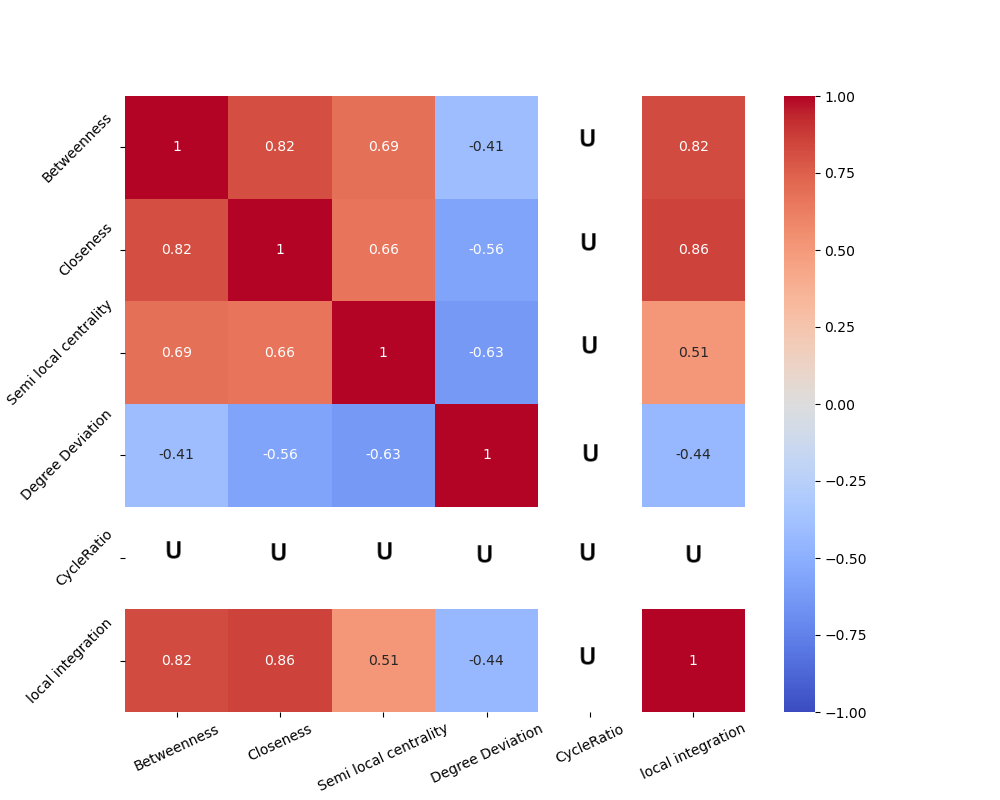}
      \caption{}
      \label{fig:subfig3}
    \end{subfigure}
    \begin{subfigure}[b]{0.45\textwidth}
      \includegraphics[width=\textwidth]{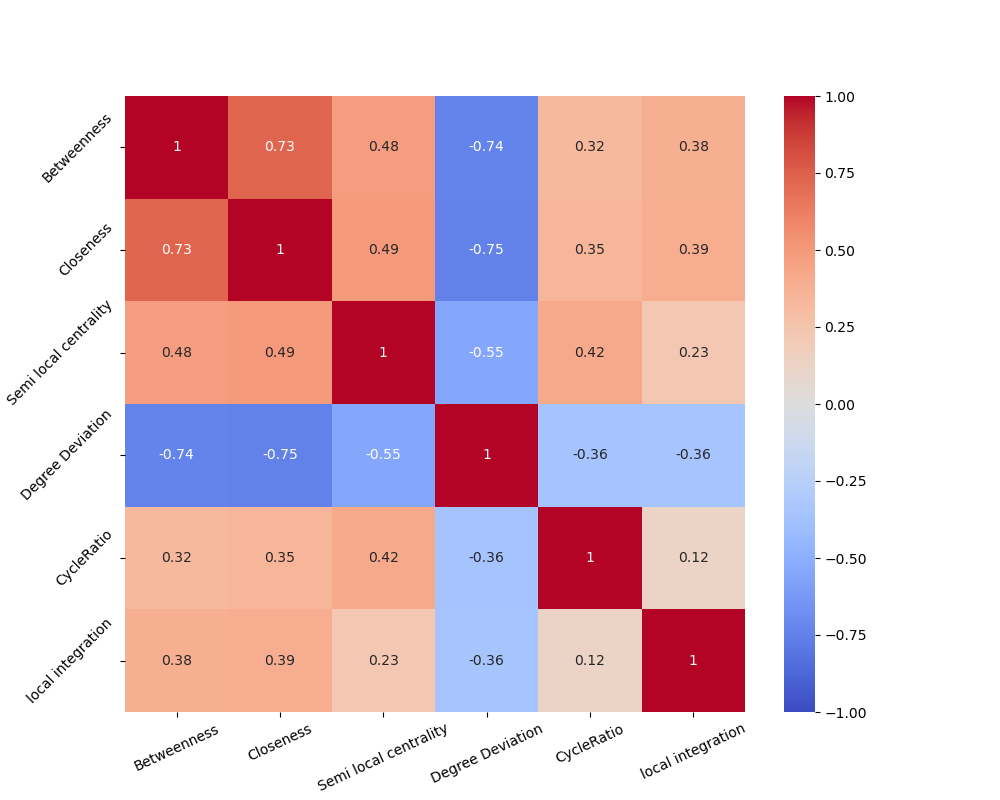}
      \caption{}
      \label{fig:subfig4}
    \end{subfigure}
    \label{fig:pair2}
  \end{subfigure}
  \caption{The Pearson correlation coefficient $\rho$ between measures in different networks: (a) high-school, (b) conference, (c) hospital, (d) workplace. ``U'' indicates undefined correlation.}
  \label{fig:correlation}
\end{figure}

We analyze the similarity between all measures discussed in this study via a Pearson correlation analysis to check if they are capturing different information for the same nodes. Highly correlated measures indicate that they are strongly similar, suggesting that one can serve as a good proxy for the other. In critical situations such as disaster handling, using one of these correlated measures ensures we do not lose much accuracy. Conversely, when analyzing a network, we can select measures with low correlation since they represent different features of the network or provide different analytical perspectives. Figure \ref{fig:correlation} shows that for all studied networks, betweenness, closeness, and semi-local centrality have the highest correlations, implying that there is no gain in using them together since they rank the nodes similarly. On the other hand, degree deviation shows an almost negative correlation with all other measures. Additionally, the cycle ratio has a low or no correlation with other measures, as seen in the hospital dataset. Therefore, in any case, the cycle ratio can be one of the selected measures.

Since there is no temporal cycle in the dataset related to the hospital, the cycle ratio for the hospital network is a constant value. As Pearson correlation relies on the variability of data points, the correlation for the cycle ratio is undefined (NaN) in the hospital dataset. This is represented by the white color in Figure \ref{fig:subfig3}

\section{Conclusions}
\label{conclusion}

Since social interactions change over time, studying and designing algorithms to characterise temporal networks is helpful. Detecting these networks' critical or more influential nodes is essential because they may be used to control epidemic outbreaks, opinions, and marketing campaigns. We introduced three novel measures for identifying the most critical nodes in temporal networks, considering both the local and global features of the nodes.

We applied these measures to four real-world contact networks in different contexts: a conference, school, workplace and hospital. The first measure, temporal supracycle ratio (TSCR), is based on the total number of cycles in which a node is involved; a node involved in more cycles is deemed more important. The second measure, temporal semi-local integration (TSLI), indicates that nodes connected to important edges, defined as edges with more active time, are also important. The third measure, temporal semi-local centrality (TSLC), is based on the second-order neighbours of a node, reflecting its semi-local centrality.

First, we ranked the nodes using these measures and compared the results with known measures, including temporal betweenness, temporal closeness, and temporal degree deviations. We analysed the accuracy of these measures by examining their effect on controlling an epidemic by removing the most critical nodes. The proposed measures demonstrated superior performance in terms of epidemic spread. By removing the critical nodes identified by these measures, the measure performs better as much as the peak value of $\Omega$ is lower. In the high school network, TSLC showed the best performance; in the conference dataset, TSCR was most effective; in the hospital dataset, both TSCR and TSLI were optimal; and in the workplace dataset, TSCR was the best for controlling the epidemic through node removal. We also removed different fractions of critical nodes to analyse their role in epidemic spread, and the proposed measures consistently performed well compared to other known measures.

Selecting the most influential nodes as the initial spreaders is crucial for opinion or information propagation, as they can accelerate the spread and affect more individuals. In our experiments, the proposed measures exhibited the highest peak values of $\Omega$ and reached $\Omega = 0$ (where the entire population is infected) sooner than others across all four networks. The robustness of networks is also significantly affected by the removal of critical nodes. The more critical the nodes, the more their removal leads to network disconnection, resulting in a smaller size of the largest connected component. The proposed measures, particularly TSCR and TSLC, generally performed better than known measures for different networks. TSLI performs similarly to betweenness but is less effective.

Different measures can be helpful depending on network features, such as density and degree deviation, as each focuses on specific features. Therefore, in situations where accuracy is crucial, it is advisable to use multiple metrics to ensure the best selection. Measures like betweenness, degree deviation, and closeness consider nodes from only one aspect, focusing on global or local features. These measures do not account for the semi-local features of nodes, such as groups of friends or coworkers. Therefore, semi-local measures show better performance in networks where people are usually network community members.

Studying the features of temporal networks aids policymakers in controlling epidemics, hindering or accelerating the spread of information, such as news or ads. The robustness of networks is necessary because removing critical nodes leads to network fragmentation, creating smaller connected components. This disconnection hinders information propagation, demonstrating the importance of maintaining network integrity. In future work, we can study networks from the perspective of network communities and compare the results of community detection algorithms with the proposed measures. We can analyse the relationship between populations of different community sizes and the proposed measures.

\bibliographystyle{elsarticle-num-names} 
\bibliography{Zara_vital}

\end{document}